\documentclass[sigconf,authorversion, nonacm]{acmart}
\usepackage{color}

\newcommand{\yixuan}[1]{{{\color{blue} \textit{(Yixuan: #1)}}}}

\AtBeginDocument{%
	\providecommand\BibTeX{{%
			\normalfont B\kern-0.5em{\scshape i\kern-0.25em b}\kern-0.8em\TeX}}}

\copyrightyear{2020}
\acmYear{2020}
\setcopyright{acmcopyright}\acmConference[BuildSys '20]{The 7th ACM International Conference on Systems for Energy-Efficient Buildings, Cities, and Transportation}{November 18--20, 2020}{Virtual Event, Japan}
\acmBooktitle{The 7th ACM International Conference on Systems for Energy-Efficient Buildings, Cities, and Transportation (BuildSys '20), November 18--20, 2020, Virtual Event, Japan}
\acmPrice{15.00}
\acmDOI{10.1145/3408308.3427617}
\acmISBN{978-1-4503-8061-4/20/11}

\usepackage{algpseudocode}
\usepackage{algorithmicx,algorithm}
\usepackage{subfigure}
\usepackage{bbding}
\usepackage{pifont}
\usepackage{wasysym}

\usepackage{balance}
\usepackage{hyperref}

\newenvironment{myitemize}{\begin{list}{$\bullet$}
		{\setlength{\topsep}{1mm}
			\setlength{\itemsep}{0.25mm}
			\setlength{\parsep}{0.25mm}
			\setlength{\itemindent}{0mm}
			\setlength{\partopsep}{0mm}
			\setlength{\labelwidth}{15mm}
			\setlength{\leftmargin}{4mm}}}{\end{list}}

\begin{document}
	
	\title{One for Many: Transfer Learning for Building HVAC Control}

	\author{Shichao Xu}
	\affiliation{%
		\institution{Northwestern University}
		\city{Evanston}
		\country{USA}}
	\email{shichaoxu2023@u.northwestern.edu}
	
	\author{Yixuan Wang}
	\affiliation{%
		\institution{Northwestern University}
		\city{Evanston}
		\country{USA}}
	\email{yixuanwang2024@u.northwestern.edu}
	
	\author{Yanzhi Wang}
	\affiliation{%
		\institution{Northeastern University}
		\city{Boston}
		\country{USA}}
	\email{yanz.wang@northeastern.edu}
	
	\author{Zheng O'Neill}
	\affiliation{%
		\institution{Texas A\&M University}
		\city{College Station}
		\country{USA}}
	\email{zoneill@tamu.edu}
	
	\author{Qi Zhu}
	\affiliation{%
		\institution{Northwestern University}
		\city{Evanston}
		\country{USA}}
	\email{qzhu@northwestern.edu}

	
	\begin{abstract}
		The design of building heating, ventilation, and air conditioning (HVAC) system is critically important, as it accounts for around half of building energy consumption and directly affects occupant comfort, productivity, and health. 
		Traditional HVAC control methods are typically based on creating explicit physical models for building thermal dynamics, which often require significant effort to develop and are difficult to achieve sufficient accuracy and efficiency for runtime building control and scalability for field implementations. Recently, deep reinforcement learning (DRL) has emerged as a promising data-driven method that provides good control performance without analyzing physical models at runtime.  
		However, a major challenge to DRL (and many other data-driven learning methods) is the long training time it takes to reach the desired performance. 
		In this work, we present a novel transfer learning based approach to overcome this challenge.   
		Our approach can effectively transfer a DRL-based HVAC controller trained for the source building to a controller for the target building with minimal effort and improved performance, by decomposing the design of neural network controller into a transferable front-end network that captures building-agnostic behavior and a back-end network that can be efficiently trained for each specific building. 
		We conducted experiments on a variety of transfer scenarios between buildings with different sizes, numbers of thermal zones, materials and layouts, air conditioner types, and ambient weather conditions. The experimental results demonstrated the effectiveness of our approach in significantly reducing the training time, energy cost, and temperature violations.
	\end{abstract}

	
	
	
	
	
	\maketitle
	
	\section{Introduction}
	
	The building stock accounts for around $40\%$ of the annual energy consumption in the United States, and nearly half of the building energy is consumed by the heating, ventilation, and air conditioning (HVAC) system~\cite{DoE}. 
	On the other hand, average Americans spend approximately $87\%$ of their time indoors~\cite{klepeis2001national}, where the operation of HVAC system has a significant impact on their comfort, productivity, and health. Thus, it is critically important to design HVAC control systems that are both energy efficient and able to maintain the desired temperature and indoor air quality for occupants. 
	
	
	In the literature, there is an extensive body of work addressing the control design of building HVAC systems~\cite{salakij2016model, maasoumy2011model, wei2015proactive, yang2020distributed}. Most of them use \textbf{model-based} approaches that create simplified physical models to capture building thermal dynamics for efficient HVAC control. For instance, resistor-capacitor (RC) networks are used for modeling building thermal dynamics in~\cite{maasoumy2011model, maasoumy2014handling, maasoumy2014selecting}, and linear-quadratic regulator (LQR) or model predictive control (MPC) based approaches are developed accordingly for efficient runtime control. 
	However, creating a simplified yet sufficiently-accurate physical model for runtime HVAC control is often difficult, as building room air temperature is complexly affected by a number of factors, including building layout, structure, construction and materials, surrounding environment (e.g., ambient temperature, humidity, and solar radiation), internal heat generation from occupants, lighting, and appliances, etc. Moreover, it takes significant effort and time to develop explicit physical models, find the right parameters, and update the models over the building lifecycle~\cite{wei2019TSUC}.    
	
	The drawbacks of model-based approaches have motivated the development of \textbf{data-driven} HVAC control methods that do not rely on analyzing physical models at runtime but rather directly making the decisions based on input data. A number of data-driven methods such as reinforcement learning (RL) have been proposed in the literature, including more traditional methods that leverage the classical Q-learning techniques and perform optimization based on a tabular $Q$ value function~\cite{barrett2015Springer, li2015CASE, nikovski2013REHVA}, earlier works that utilize neural networks~\cite{fazenda2014AISE, costanzo2016experimental}, and more recent deep reinforcement learning (DRL) methods~\cite{wei2017deep, zhang2018practical, zhang2018deep, li2019transforming, naug2019online,  gao2019energy, gao2020deepcomfort}. In particular, the DRL-based methods leverage deep neural networks for estimating the $Q$ values associated with state-action pairs and are able to handle larger state space than traditional RL methods~\cite{wei2019TSUC}. They have emerged as a promising solution that offers good HVAC control performance without analyzing physical models at runtime.
	
	
	
	However, there are major challenges in deploying DRL-based methods in practice. Given the complexity of modern buildings, it could take a significant amount of training for DRL models to reach the desired performance. For instance, around 50 to 100 months of data are needed for training the models in~\cite{wei2017deep, wei2019TSUC} and 4000+ months of data are used for more complex models~\cite{yu2020multi, gao2020deepcomfort} -- even if this could be drastically reduced to a few months or weeks, directly deploying DRL models on operational buildings and taking so long before getting the desired performance is impractical. The works in~\cite{wei2017deep, wei2019TSUC} thus propose to first use detailed and accurate physical models (e.g., EnergyPlus~\cite{Crawley00energyplus:energy}) for offline simulation-based training before the deployment. While such an approach can speed up the training process, it still requires the development and update of detailed physical models, which as stated above needs significant domain expertise, effort, and time.
	
	
	To address the challenges in DRL training for HVAC control, we propose a \textbf{transfer learning} based approach in this paper, to utilize existing models (that had been trained for old buildings) in the development of DRL methods for new buildings.  
	This is not a straightforward process, however. Different buildings may have different sizes, numbers of thermal zones, materials and layouts, HVAC equipment, and operate under different ambient weather conditions. As shown later in the experiments, directly transferring models between such different buildings is not effective. 
	In the literature, there are a few works that have explored transfer learning for buildings.
	In~\cite{chen2020transfer}, a building temperature and humidity prediction model is learned from supervised learning, and transferred to new buildings with further tuning and utilized in an MPC algorithm. The work in~\cite{lissa2020transfer} investigates the transfer of Q-learning for building HVAC control under different weather conditions and with different room sizes, but it is limited to single-room buildings. The usage of Q-table in conventional Q-learning also leads to limited memory for state-action pairs and makes it unsuitable for complex buildings. 
	
	
	Our work addresses the limitations in the literature, and develops for the first time a Deep Q-Network (DQN) based transfer learning approach for multiple-zone buildings. Our approach avoids the development of physical models, significantly reduces the DRL training time via transfer learning, and is able to reduce energy cost while maintaining room temperatures within desired bounds. More specifically, our work makes the following contributions:
	\begin{myitemize}
		\item We propose a novel transfer learning approach that decomposes the design of neural network based HVAC controller into two (sub-)networks. The front-end network captures building-agnostic behavior and can be directly transferred, while the back-end network can be efficiently trained for each specific building in an offline supervised manner by leveraging a small amount of data from existing controllers (e.g., simple on-off controller). 
		\item Our approach requires little to no further tuning of the transferred DRL model after it is deployed in the new building, thanks to the two-subnetwork design and the offline supervised training of the back-end network. This avoids the initial \emph{cold start} period where the HVAC control may be unstable and unpredictable.
		\item We have performed a number of experiments for evaluating the effectiveness of our approach under various scenarios. The results demonstrate that our approach can effectively transfer between buildings with different sizes, numbers of thermal zones, materials and layouts, and HVAC equipment, as well as under different weather conditions in certain cases. Our approach could enable fast deployment of DRL-based HVAC control with little training time after transfer, and reduce building energy cost with minimal violation of temperature constraints.   
	\end{myitemize}


	
	
	The rest of the paper is structured as follows. Section~\ref{related_work} provides a more detailed review of related work. 
	Section~\ref{methodology} presents our approach, including the design of two networks and the corresponding training methods.
	Section~\ref{experiment} shows the experiments for different transfer scenarios and other related ablation studies. Section~\ref{conclusion} concludes the paper.
	
	\section{Related work}
	\label{related_work}
	
	\noindent
	\textbf{Model-based and Data-driven HVAC Control.} There is a rich literature in HVAC control design, where the approaches can generally fall into two main categories, i.e., model-based and data-driven.
	
	
	Traditional model-based HVAC control approaches typically build explicit physical models for the controlled buildings and their surrounding environment, and then design control algorithms accordingly~\cite{salakij2016model, maasoumy2011model}. For instance, the work in~\cite{ma2012TCST} presents a nonlinear model for the overall cooling system, which includes chillers, cooling towers and thermal storage tanks, and then develops an MPC-based approach for reducing building energy consumption. The work in~\cite{maasoumy2011model} models the building thermal dynamics as RC networks, calibrates the model based on historical data, and then presents a tracking LQR approach for HVAC control. Similar simplified models have been utilized in other works~\cite{maasoumy2014selecting, maasoumy2014handling, wei2015proactive} for HVAC control and for co-scheduling HVAC operation with other energy demands and power supplies. While being efficient, these simplified models often do not provide sufficient accuracy for effective runtime control, given the complex relation between building room air temperature and various factors of the building itself (e.g., layout, structure, construction and materials), its surrounding environment (e.g., ambient temperature, humidity, solar radiation), and internal operation (e.g., heat generation from occupants, lighting and appliances). More accurate physical models can be built and simulated with tools such as EnergyPlus~\cite{Crawley00energyplus:energy}, but those models are typically too complex to be used for runtime control.

	
	Data-driven approaches have thus emerged in recent years due to their advantages of not requiring explicit physical models at runtime
	. These approaches often leverage various machine learning techniques, in particular reinforcement learning. 
	For instance, in~\cite{wei2017deep, zhang2018practical}, DRL is applied to building HVAC control and an EnergyPlus model is leveraged for simulation-based offline training of DRL. In~\cite{zhang2018deep, gao2019energy}, DRL approaches leveraging the actor-critic methods are applied. The works in~\cite{gao2020deepcomfort, naug2019online} use data-driven methods to approximate/learn the energy consumption and occupants' satisfaction under different thermal conditions, and then apply DRL to learn an end-to-end HVAC control policy.  
	These DRL-based methods are shown to be effective at reducing energy cost and maintaining desired temperature, and are sufficiently efficient at runtime. However, they often take a long training time to reach the desired performance, needing dozens and hundreds of months of data for training~\cite{wei2017deep, wei2019TSUC} or even longer~\cite{gao2020deepcomfort, yu2020multi}. Directly deploying them in real buildings for such long training process is obviously not practical. Leveraging tools such as EnergyPlus for offline simulation-based training can mitigate this issue, but again incurs the need for the expensive and sometimes error-prone process of developing accurate physical models (needed for simulation in this case). These challenges have motivated this work to develop a transfer learning approach for efficient and effective DRL control of HVAC systems.
	
	\smallskip
	\noindent
	\textbf{Transfer Learning for HVAC control.} There are a few works that have explored transfer learning in buildings HVAC control. In~\cite{lissa2020transfer}, transfer learning of a Q-learning agent is studied, however only a single room (thermal zone) is considered. The usage of a tabular table for each state-action pair in the traditional Q-learning in fact limits the approach's capability to handle high-dimensional data. In~\cite{chen2020transfer}, a neural network model for predicting temperature and humidity is learned in a supervised manner and transferred to new buildings for MPC-based control. The approach also focuses on single-zone buildings and requires further tuning after the deployment of the controller.
	
	Different from these earlier works in transfer learning for HVAC control, our approach addresses multi-zone buildings and considers transfer between buildings with different sizes, number of thermal zones, layouts and materials, HVAC equipment, and ambient weather conditions. It also requires little to no further tuning after the transfer. This is achieved with a novel DRL controller design with two sub-networks and the corresponding training methods.

	
	
	
	\smallskip
	\noindent
	\textbf{Transfer Learning in DRL.} Since our approach considers transfer learning for DRL, it is worth to note some of the work in DRL-based transfer learning for other domains~\cite{zhan2015online, gupta2017learning, da2019survey, akkaya2019solving}. For instance, in~\cite{gupta2017learning}, the distribution of optimal trajectories across similar robots is matched for transfer learning in robotics. In~\cite{akkaya2019solving}, an environment randomization approach is proposed, where DRL agents trained in simulation with a large number of generated environments can be successfully transferred to their real-world applications.
	To the best of our knowledge, our work is the first to propose DRL-based transfer learning for multi-zone building HVAC control. It addresses the unique challenges in building domain, e.g., designing a novel two-subnetwork controller to avoid the complexity and cost of creating accurate physical models for simulation.

	\section{Our Approach}
	\label{methodology}
	\begin{figure*}[htbp]
		\centering
		\setlength{\abovecaptionskip}{0.15cm}
		\setlength{\belowcaptionskip}{-0.25cm}
		\includegraphics[width=12.2cm]{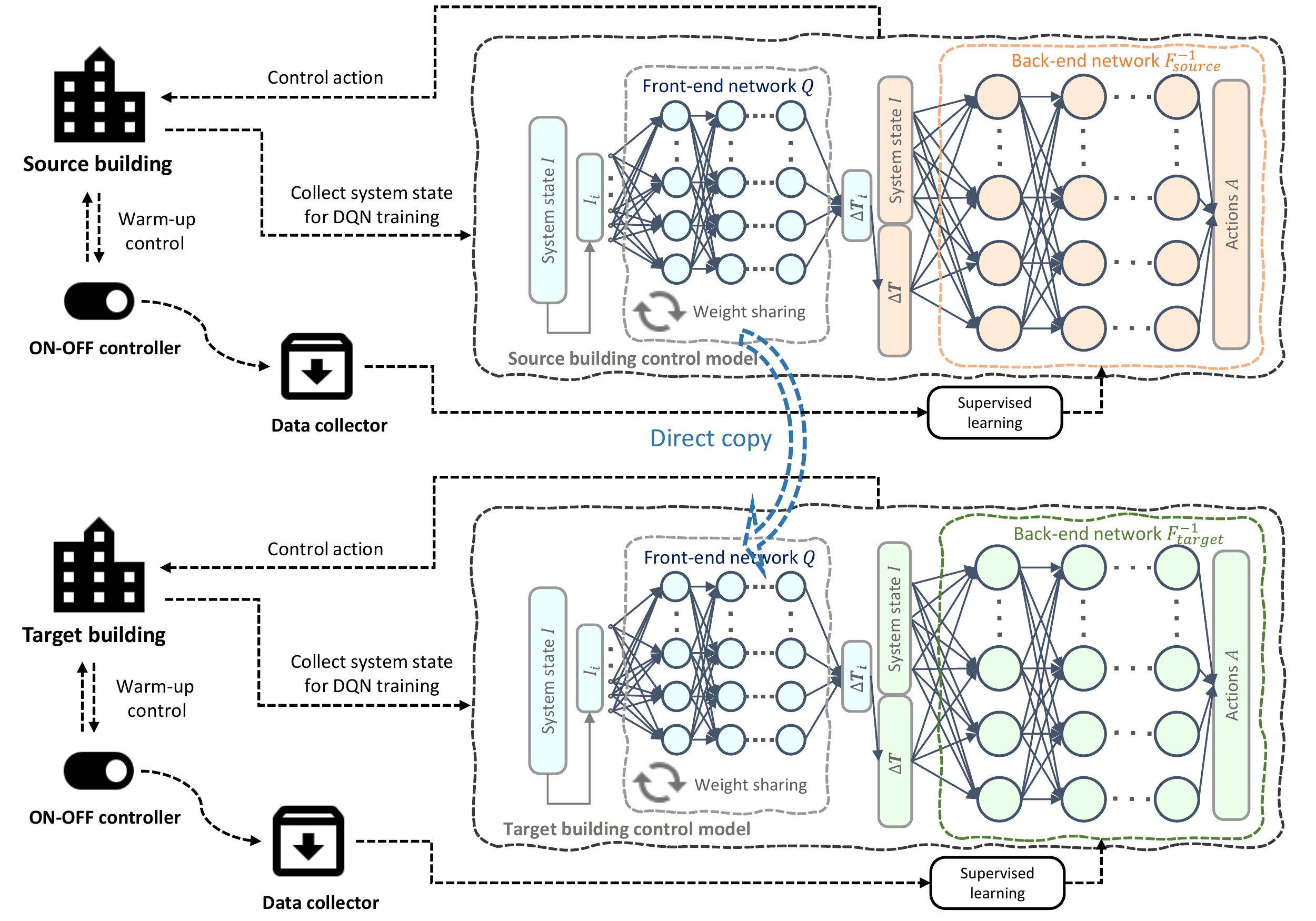}
		\caption{\small Overview of our DRL-based transfer learning approach for HVAC control.  
			We design a novel DQN architecture that includes two sub-networks: A \textbf{front-end network} $Q$ captures the building-agnostic part of the control as much as possible, while a \textbf{back-end network} (inverse building network) $F^{-1}$ captures the building-specific behavior. 
			At each control step, the front-end network $Q$ maps the current system state $I$ to an intermediate state $\Delta T$. Then, the back-end network $F^{-1}$ maps $\Delta T$, together with $I$, to the control action outputs $\mathbf{A}$.
			During transfer learning from a source building to a target building, the front-end network $Q$ is directly transferable. The back-end network $F^{-1}$ can be trained in a supervised manner, with data collected from an existing controller (e.g., a simple ON-OFF controller). Experiments have shown that around two weeks of data is sufficient for such supervised training of $F^{-1}$. If it is a brand new building without any existing controller, we can deploy a simple ON-OFF controller for two weeks in a ``warm-up'' process. During this process, the ON-OFF controller can maintain the temperature within the desired bounds (albeit with higher cost), and collect data that captures the building-specific behavior for training $F^{-1}$.}
		\label{overview}
	\end{figure*}
	
	We present our transfer learning approach in this section, including the design of the two-subnetwork controller and the training process. Section~\ref{sec3-1} introduces the system model. Section~\ref{sec3-2} provides an overview of our methodology. Section~\ref{sec3-3} presents the design of the building-agnostic front-end (sub-)network, and Section~\ref{sec3-4} explains the design of the building-specific back-end (sub-)network.
	

	\subsection{System Model}
	\label{sec3-1}
	The goal of our work is to build a transferable HVAC control system that can maintain comfortable room air temperature within desired bounds while reducing the energy cost. 
	We adopt a building model that is similar to the one used in~\cite{wei2017deep}, an $n$-zone building model with a variable air volume (VAV) HVAC system. The system provides conditioned air at a flow rate chosen from $m$ discrete levels. Thus, the entire action space for the $n$-zone controller can be described as $\mathbf{A} = \{\mathbf{a_1, a_2, \cdots, a_n}\},$
	where $\mathbf{a_i} (1 \leq i \leq n)$ is chosen from $m$ VAV levels $\{f_1, f_2, \cdots, f_m\}$. Note that the size of the action space ($m^n$) increases exponentially with respect to the number of thermal zones $n$, which presents significant challenge to DRL control for larger buildings. 
	We address this challenge in the design of our two-subnetwork DRL controller by avoiding setting the size of the neural network action output layer to $m^n$.  This will be explained further later.  
	
	The DRL action is determined by the current system state. In our model, the system state includes the current physical time $t$, inside state $S_{in}$, and outside environment state $S_{out}$. The inside state $S_{in}$ includes the temperature of each thermal zone, denoted as $\{T_{1}, T_{2}, \cdots, T_{n}\}$. The outside environment state $S_{out}$ includes the ambient temperature and the solar irradiance (radiation intensity). Similar to~\cite{wei2017deep}, to improve DRL performance, $S_{out}$ not only includes the current values of the ambient temperature $T_{out}^i$ and the solar irradiance $Sun_{out}^i$, but also their weather forecast values for the next three days.
	Thus, the outside environment state is denoted as $S_{out} = \{T_{out}^0, T_{out}^1, T_{out}^2, T_{out}^3, Sun_{out}^0, Sun_{out}^1, Sun_{out}^2, Sun_{out}^3\}$. 
	Our current model does not consider internal heat generation from occupants, a limitation that we plan to address in future work. 
	
	
	\subsection{Methodology Overview}
	\label{sec3-2}
	
	We started our work by considering whether it is possible to directly transfer a well-trained DQN model for a single-zone source building to every zone of a target multiple-zone building. However, based on our experiments (shown later in Table~\ref{result1-2} of Section~\ref{experiment}), such straightforward approach is not effective at all, leading to significant temperature violations. 
	This is perhaps not surprising. In DQN-based reinforcement learning, a neural network $Q$ maps the input $I = \{I_1, I_2, \cdots, I_n\}$, where $I_i$ is the state for each zone $i$, to the control action output $\mathbf{A}$. The network $Q$ is optimized based on a reward function that considers energy cost and temperature violation. Through training, $Q$ learns a control strategy that incorporates the consideration of building thermal dynamics, including the building-specific characteristics. Directly applying $Q$ to a new target building, which may have totally different characteristics and dynamics, will not be effective in general.  
	
	
	Thus, our approach designs a novel architecture that includes two sub-networks, with an intermediate state $\Delta T$ that indicates a predictive value of the controller's willingness to change the indoor temperature. The \textbf{front-end network} $Q$ maps the inputs $I$ to the intermediate state $\Delta T$. It is trained to capture the building-agnostic part of the control strategy, and is directly transferable. The \textbf{back-end network} then maps $\Delta T$, together with $I$, to the control action output $\mathbf{A}$. It is trained to capture the building-specific part of the control, and can be viewed as an inverse building network $F^{-1}$. An overview of our approach is illustrated in Figure~\ref{overview}.
	
	
	\subsection{Front-end Building-agnostic Network Design and Training
	}
	\label{sec3-3}
	We introduce the design of our front-end network $Q$ and its training in this section. 
	$Q$ is composed of $n$ (sub-)networks itself, where 
	where $n$ is the number of building thermal zones. Each zone in the building model has its corresponding sub-network, and all sub-networks share their weights. In each sub-network for thermal zone $i$, the input layer accepts state $I_i$. It is followed by $L$ sequentially-connected fully-connected layers (the exact number of neurons is presented later in Table~\ref{hyperparam} of Section~\ref{experiment}). Rather than directly giving the control action likelihood vector, the network's output layer reflects a planned temperature change value $\Delta T_i$ for each zone.

	More specifically, the output of the last layer is designed as a vector $O_{\Delta T_i}$ of length $h+2$ in one-hot representation -- 
	the planned temperature changing range is equally divided into $h$ intervals within a predefined temperature range of $[-b, b]$ and two intervals outside of that range are also considered. The relationship of the planned temperature change value $\Delta T_i$ of zone $i$ and the output vector $O_{\Delta T_i}$ is as follows:
	
	\begin{equation}
	O_{\Delta T_i} = \left\{
	\begin{array}{lr}
	<1, 0, \cdots, 0>, &  \Delta T_i \leq -b,\\
	< 0, \cdots, 0,1,0,\cdots, 0>, &  -b < \Delta T_i < b,\\
	\scriptstyle{the\ position\ of\ 1\ is\ at\ (\lfloor \Delta T_i / (2b / h) \rfloor))}&\\
	<0, \cdots, 0, 1>, & \Delta T_i \geq b.
	\end{array}
	\right.
	\end{equation}
	
	Then, for the entire front-end network $Q$, the combined input is $I = \{I_1, I_2, \cdots, I_n\}$, and the combined output is $O_{\Delta T} = \{O_{\Delta T_1}, O_{\Delta T_2}, \cdots, O_{\Delta T_n}\}.$
	
	
	It is worth noting that if we had designed the front-end network in standard deep Q-learning model~\cite{mnih2015human}, it would take $I$ as the network's input, pass it through several fully-connected layers, and output the selection among an action space that has a size of $(h+2)^n$ (as there are $n$ zones, and each has $h+2$ possible actions). It also needs an equal number of neurons for the last layer, which is not affordable when the number of zones gets large. Instead in our design, the last layer of the front-end network $Q$ has its size reduced to $(h+2)*n$, which can be further reduced to $(h+2)$ with the following weight-sharing technique.  
	
	We decide to let the $n$ sub-networks of $Q$ share their weights during training. One benefit of this design is that it enables transferring the front-end network for a $n$-zone source building to a target $m$-zone building, where $m$ could be different from $n$. It also reduces the training load by lowering the number of parameters. Such design performs well in our experiments.

	Our front-end network $Q$ is trained with the standard deep Q-learning techniques~\cite{mnih2015human}. Note that while the output action for $Q$ is the planned temperature change vector $O_{\Delta T}$, the training process uses a dynamic reward $R_t$ that depends on the eventual action (i.e., output of network $F^{-1}$), which will be introduced later in Section~\ref{sec3-4}.
	Specifically, the training of the front-end network $Q$ follows Algorithm~\ref{algo1} (the hyper-parameters used are listed later in Table~\ref{hyperparam} of Section~\ref{experiment}). 
	First, we initialize $Q$ by following the weights initialization method described in ~\cite{he2015delving} and copy its weights to the target network $Q^{'}$ (target network $Q^{'}$ is a technique in deep Q-learning that is used for improving performance.). The back-end network $F^{-1}$ is initialized following Algorithm~\ref{algo2} (introduced later in Section~\ref{sec3-4}). We also empty the replay buffer and set the exploration rate $\epsilon$ to 1.  
	
	
	
	At each control instant $t$ during a training epoch, we obtain the current system state $S_{cur} =$ ($t$, $S_{in}$, $S_{out}$) and calculate the current reward $R_t$. We then collect the learning samples (experience) ($S_{pre}$, $S_{cur}$, $\Delta T$, $\mathbf{A}$, $R$) and store them in the replay buffer.
	In the following learning-related operations, we first sample a data batch $M = (\mathscr{S}_{prime}, \mathscr{S}_{next}, \mathscr{a}, \mathscr{r})$ from the replay buffer, and calculate the actual temperature change value $\Delta \mathscr{T}_a $ from $\mathscr{S}_{prime}$ and $ \mathscr{S}_{next}$. Then, we get the planned temperature change value from the back-end network $F^{-1}$, i.e., $\mathscr{a}_p$ = $F^{-1}(\Delta \mathscr{T}_a, \mathscr{S}_{prime})$. In this way, the cross entropy loss can be calculated from the true label $\mathscr{a}$ and the predicted label $\mathscr{a}_p$. We then use supervised learning to update the back-end network $F^{-1}$ with the Adam optimizer~\cite{kingma2014adam} under learning rate $lr_2$. 
	
	We follow the same procedure as described in~\cite{mnih2015human} to calculate the target vector $v$ that is used in deep Q-learning.
	With target vector $v$ and input state $S_{prime}$, we can then train $Q$ using the back-propagation method~\cite{goodfellow20166} with mean squared error loss and learning rate $lr_1$. With a period of $\Delta nt$, we assign the weights of $Q$ to the target network $Q^{'}$. The exploration rate is updated as $\epsilon = \max\{\epsilon_{low}, \epsilon-\Delta \epsilon\}$. It is used for $\epsilon-$greedy policy to select each planned temperature change value $\Delta T_i$:
	\begin{equation}
	\Delta T_i = \left.\{ \begin{aligned}
	& argmax\ O_{\Delta T_i}\ & with\ probability\ & 1-\epsilon,\\
	& random(0\ to\ h+1)\ & with\ probability\ & \epsilon.\\
	\end{aligned}\right.\\
	\end{equation}
	
	\begin{equation}
	\Delta T = \{\Delta T_1, \Delta T_2, \cdots, \Delta T_n\}.
	\end{equation}
	The control action $\mathbf{A}$ is obtained from the back-end network:
	\begin{equation}
	\mathbf{A} = F^{-1}(\Delta T, S_{cur}).
	\end{equation}
	
	\begin{algorithm}[t]
		\small
		
		\caption{Training of front-end network $Q$}
		\label{algo1}
		
		\begin{algorithmic}[1]
			\State $ep$: the number of training epochs
			\State $\Delta ct$: the control period
			\State $t_{MAX}$: the maximum training time of an epoch
			\State $\Delta nt$: the time interval to update target network
			\State Empty replay buffer
			\State Initialize $Q$; set the weights of target network $Q^{'}$ = $Q$; initialize $F^{-1}$ based on Algorithm~\ref{algo2}
			\State Initialize the current planned temperature change vector ${\Delta T}$
			
			\State Initialize previous state $S_{pre}$
			\State Initialize exploration rate $\epsilon$
			
			\For{$Epoch$ = 1 to $ep$}
			\For{$t$ = 0 to $t_{MAX}$, $t$ += $\Delta ct$}
			\State $S_{cur} \leftarrow$ ($t$, $S_{in}$, $S_{out}$)
			
			\State Calculate reward $R$
			\State Add experience ($S_{pre}$, $S_{cur}$, $\Delta T$, $\mathbf{A}$, $R$) to the replay buffer
			\For{$tr$ = 0 to $L_{MAX}$}
			\State Sample a batch $M = (\mathscr{S}_{prime}, \mathscr{S}_{next}, \mathscr{a}, \mathscr{r})$
			
			\State Calculate actual temperature change value $\Delta \mathscr{T}_a$  
			
			\State Predicted label $\mathscr{a}_p$ = $F^{-1}(\Delta \mathscr{T}_a, \mathscr{S}_{prime})$
			\State Set loss $L = CrossEntropyLoss(\mathscr{a}_p, \mathscr{a})$
			\State Update $F^{-1}$ with loss $L$ and learning rate $lr_2$

			\State Target $\mathscr{v}$ $\leftarrow$ target network $Q^{'}(\mathscr{S}_{prime})$
			\State Train network $Q$ with $\mathscr{S}_{prime}$ and  $\mathscr{v}$
			\EndFor
			\If {$t$ mod $\Delta nt$ == 0}
			\State Update target network $Q^{'}$
			\EndIf

			\State $O_{\Delta T} = Q(S_{cur})$
			\State Update exploration rate $\epsilon$
			\State Update each $\Delta T_i$ follows $\epsilon-$greedy policy
			\State $\Delta T = <\Delta T_1, \Delta T_2, \cdots, \Delta T_n>$
			\State Control action $\mathbf{A} \leftarrow$ $F^{-1}(\Delta T, S_{cur})$
			\State $S_{pre} = S_{cur}$
			
			\EndFor
			\EndFor
			
		\end{algorithmic}
	\end{algorithm}

	\subsection{Back-end Building-specific Network Design and Training} 
	\label{sec3-4}
	
	The objective of the back-end network is to map the planned temperature change vector $O_{\Delta T}$ (or $\Delta T$), together with the system state $I$, into the control action $\mathbf{A}$.
	Consider that during operation, a building environment ``maps'' the control action and system state to the actual temperature change value. So in a way, the back-end network can be viewed as doing the \emph{inverse} of what a building environment does, i.e., it can be viewed as an inverse building network $F^{-1}$. 
	
	The network $F^{-1}$ receives the planned temperature change value $\Delta T$ and the system state $I$ at its input layer. It is followed by $L^{'}$ fully-connected layers (exact number for experimentation is specified in Table~\ref{hyperparam} of Section~\ref{experiment}). 
	It outputs a likelihood control action vector $O_{\mathbf{A}} = \{v_1, v_2, \cdots, v_n\}$, which can be divided into $n$ groups. For group $i$, it has a one-hot vector $v_i$ corresponding to the control action for zone $i$. The length of $v_i$ is $m$, as there are $m$ possible control actions for each zone as defined earlier.
	When $O_{\mathbf{A}}$ is provided, control action $\mathbf{A}$ can be easily calculated by applying argmax operation for each group in $O_{\mathbf{A}}$, i.e., $\mathbf{A} = \{argmax\{v_1\}, argmax\{v_2\}, \cdots, argmax\{v_n\}\}$.
	
	The network $F^{-1}$ is integrated with the reward function $R_t$: 
	{
		\setlength\abovedisplayskip{0pt}
		\setlength\belowdisplayskip{0pt}
		\begin{equation}
		\label{eq3}
		R_t = {w}_{cost} {R\_cost}_t + {w}_{vio} {R\_vio}_t,
		\end{equation}
	}
	where ${R\_cost}_t$ is the reward of energy cost at time step $t$ and ${w}_{cost}$ is the corresponding scaling factor. ${R\_vio}_t$ is the reward of zone temperature violation at time step $t$ and ${w}_{vio}$ is its scaling factor. The two rewards are further defined as:
	
	\begin{equation}
	\begin{aligned}
	{R\_cost}_t = &- cost(F^{-1}({\Delta T}_{t-1}), t-1).
	\end{aligned}
	\end{equation}
	{
		\setlength\abovedisplayskip{0pt}
		\setlength\belowdisplayskip{0pt}
		\begin{equation}
		{R\_vio}_t = -\sum_{i=1}^{n}{max(T^i_t - {T_{upper}}, 0) + max({T_{lower}} - T^i_t, 0)}.
		\end{equation}
	}
	Here, $cost(,)$ is a function that calculates the energy cost within a control period according to the local electricity price that changes over time. ${\Delta T}_{t-1}$ is the planned temperature change value at time $t-1$. $T^i_t$ is the zone $i$ temperature at time $t$. ${T_{upper}}$ and ${T_{lower}}$ are the comfortable temperature upper / lower bound, respectively.
	
	As stated before, $F^{-1}$ can be trained in a supervised manner. We could also directly deploy our DRL controller, with transferred front-end network $Q$ and an initially-randomized back-end network $F^{-1}$; but we have found that leveraging data collected from the existing controller of the target building for offline supervise learning of $F^{-1}$ before deployment can provide significantly better results than starting with a random $F^{-1}$. This is because that the data from the existing controller provides insights into the building-specific behavior, which after all is what $F^{-1}$ is for. In our experiments, we have found that a simple existing controller such as the ON-OFF controller with two weeks of data can already be very effective for helping training $F^{-1}$. Note that such supervised training of $F^{-1}$ does not require the front-end network $Q$, which means $F^{-1}$ could be well-trained and ready for use before $Q$ is trained and transferred. In the case that the target building is brand new and there is no existing controller, we can deploy a simple ON-OFF controller for collecting such data in a warm-up process (Figure~\ref{overview}). While such ON-OFF controller typically consumes significantly higher energy, it can effectively maintain the room temperature within desired bounds, which means that the building could already be in use during this period. Once $F^{-1}$ is trained, the DRL controller can replace the ON-OFF controller in operation. 
	
	
	Algorithm~\ref{algo2} shows the detailed process for the training of $F^{-1}$. Note that the initialization of $F^{-1}$ in this algorithm also follows the weights initialization method described in~\cite{he2015delving}. We also augment the collected training data to ensure the boundary condition. The augmenting data is created by copying all samples from the collected data and set temperature change value $\Delta \mathscr{T}$ to the lowest level ($<-b$) while setting all control actions to the maximum level. 
	
	\begin{algorithm}[t]
		\small
		\caption{Training of back-end network $F^{-1}$}
		\label{algo2}
		
		\begin{algorithmic}[1]
			\State $ep_F$: the number of training epochs
			\State $\Delta ct$: the control period
			\State $t_{MAX}^{'}$: the maximum data collection time
			\State Initialize previous state $S_{pre}$
			\State Initialize $F^{-1}$
			\State Empty database $M$ and dataset $D$
			
			\For{$t$ = 0 to $t_{MAX}$, $t$ += $\Delta ct$}
			\State $S_{cur} \leftarrow$ ($t$, $S_{in}$, $S_{our}$)
			\State Control action $\mathbf{A}$ $\leftarrow$ run ON-OFF controller on $S_{cur}$
			\State $S_{pre} = S_{cur}$
			\State Add sample $(S_{cur}, S_{pre}, \mathbf{A})$ to database $M$
			\EndFor
			
			\For{each sample $\mathbf{u}$=($S_{cur}, S_{pre}, \mathbf{a}$) in $M$}
			
			\State $\Delta \mathscr{T}_a \leftarrow$ calculate temperature difference in ($\mathscr{S}_{cur}, \mathscr{S}_{pre}$)
			\State Add sample $\mathbf{v} = (\Delta \mathscr{T}_a, S_{pre}, \mathbf{a})$ to dataset $D$
			\EndFor
			
			\For{each sample $\mathbf{u}$=($S_{cur}, S_{pre}, \mathbf{a}$) in $M$}
			
			\State $\Delta \mathscr{T}_a \leftarrow$ lowest level
			\State $\mathbf{a^{'}} \leftarrow$ maximum air condition level
			\State Add sample $\mathbf{v} = (\Delta \mathscr{T}_a, S_{pre}, \mathbf{a^{'}})$ to dataset $D$
			\EndFor
			\For{$Epoch$ = 1 to $ep_F$}
			\For{each training batch of $(\Delta \mathscr{T}_a, S_{pre}, \mathbf{a})$ in dataset $D$}
			\State network inputs = $(\Delta \mathscr{T}_a, S_{pre})$
			\State corresponding labels = $(\mathbf{a})$
			\State Train network $F^{-1}$
			\EndFor
			\EndFor
			\State Return $F^{-1}$
		\end{algorithmic}
	\end{algorithm}

	\begin{algorithm}[t]
		\small
		\caption{Running of our proposed approach}
		\label{algo3}
		
		\begin{algorithmic}[1]
			\State $\Delta ct$: the control period
			\State $t_{MAX}$: the maximum testing time
			\State Initialize the weights of $Q$ with the front-end network transferred from the source building (see Figure~\ref{overview})
			\State Initialize the weights of $F^{-1}$ with weights learned using Algorithm~\ref{algo2}
			
			\For{$t$ = 0 to $t_{MAX}$, $t$ += $\Delta ct$}
			\State $S_{cur} \leftarrow$ ($t$, $S_{in}$, $S_{out}$)
			\State $\Delta T \leftarrow argmax\ Q(S_{cur})$
			\State Control action $\mathbf{A} \leftarrow$ $F^{-1}(\Delta T, S_{cur})$
			
			\EndFor

		\end{algorithmic}
	\end{algorithm}

	\medskip
	\noindent
	Once the front-end network $Q$ is trained as in Algorithm~\ref{algo1} and the back-end network $F^{-1}$ is trained as in Algorithm~\ref{algo2}, our transferred DRL controller is ready to be deployed and can operate as described in Algorithm~\ref{algo3}. Note that we could further fine-tune our DRL controller during the operation. This can be done by enabling a fine-tuning procedure that is similar to Algorithm~\ref{algo1}. The difference is that instead of initializing the Q-network $Q$ using~\cite{he2015delving}, we copy transferred Q-network weights from the source building to the target building's front-end network $Q$ and its corresponding target network $Q^{'}$. And we set $\epsilon = 0$, $\epsilon_{low} = 0$, and $L_{MAX}$ to $3$ instead of $1$. Other operations remain the same as in Algorithm~\ref{algo1}.
	
	
	
	
	\section{Experimental Results}
	\label{experiment}
	\subsection{Experiment Settings}
	
	All experiments are conducted on a server equipped with a 2.10GHz CPU (Intel Xeon(R) Gold 6130), 64GB RAM, and an NVIDIA TITAN RTX GPU card. The learning algorithms are implemented in the PyTorch learning framework. The Adam optimizer~\cite{kingma2014adam} is used to optimize both front-end networks and back-end networks. The DRL hyper-parameter settings are shown in Table~\ref{hyperparam}. In addition, to accurately evaluate our approach, we leverage the building simulation tool EnergyPlus~\cite{Crawley00energyplus:energy}. Note that EnergyPlus here is only used for evaluation purpose, in place of real buildings. During the practical application of our approach, EnergyPlus is not needed. This is different from some of the approaches in the literature~\cite{wei2017deep, wei2019TSUC}, where EnergyPlus is needed for offline training before deployment and hence accurate and expensive physical models have to be developed.

	\begin{table}
		\newcommand{\tabincell}[2]{\begin{tabular}{@{}#1@{}}#2\end{tabular}}
		\small
		\vspace{-0.0cm}
		\setlength{\abovecaptionskip}{-0.5cm}
		\setlength{\belowcaptionskip}{0.1cm}
		\begin{tabular}{c|c|c|c}  
			\hline  
			Parameter & Value & Parameter & Value  \\  
			\hline  
			\tabincell{c}{Front-end \\network layers} & \tabincell{c}{[10,128,256,\\256,256,400,22]}  & \tabincell{c}{Back-end \\network layers} & \tabincell{c}{[22*n,128,256,\\256,128,m*n]} \\
			$b$ & 2 & $h$ & 20 \\  
			$lr_1$ & 0.0003 & $ep$ & 150 \\
			$lr_2$ & 0.0001 & $ep_F$ & 15 \\
			$L_{MAX}$ & 1 & $w_{cost}$ & $\frac{1}{1000}$\\
			$ep$ & 150 & $w_{vio}$ & $\frac{1}{1600}$\\
			$T_{lower}$ & 19 & $T_{upper}$ & 24 \\
			$\Delta nt$ & 240*15 min & $\Delta ct$ & 15 min \\
			$t_{MAX}^{'}$ & 2 weeks & $t_{MAX}$ & 1 month \\
			$\epsilon_{low}$ & 0.1 &  & \\
			\hline  
		\end{tabular}  
		
		\caption{Hyper-parameters used in our experiments.}
		\label{hyperparam}
	\end{table}  
	
\begin{figure*}[htbp]
		\centering
		\vspace{-0.0cm}
		\setlength{\abovecaptionskip}{-0.0cm}
		\setlength{\belowcaptionskip}{-0.2cm}
		\includegraphics[width=18cm]{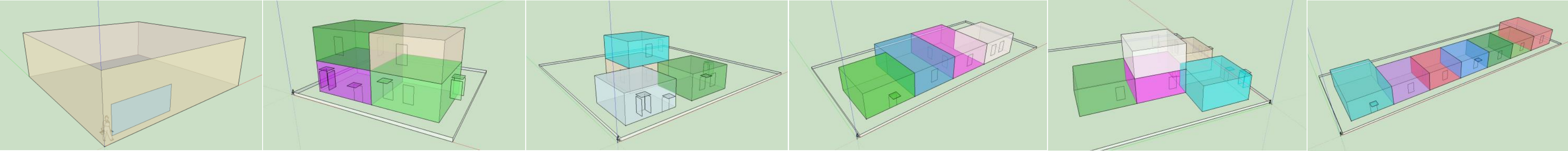}
		\caption{Different building models used in our experiments. From left to right, the models are one-zone building 1, four-zone building 1, four-zone building 2 , four-zone building 3, five-zone building 1, seven-zone building 1. Compared to four-zone building 1, four-zone building 2 has different layout and wall material; four-zone building 3 has different layout, wall material, and room size; five-zone building 1 has different number of zones, layout, and wall material; and seven-zone building 1 has different number of zones, layout, wall material, and room size.}
		\label{building visual}
	\end{figure*}

	In our experiments, simulation models in EnergyPlus interact with the learning algorithms written in Python through the Building Controls Virtual Test Bed (BCVTB)~\cite{wetter2011co}. 
	We simulate the building models with the weather data obtained from the Typical Meteorological Year 3 database~\cite{wilcox2008users}, and choose the summer weather data in August (each training epoch contains one-month data). Apart from the weather transferring experiments, all other experiments are based on the weather data collected in Riverside, California, where the ambient weather changes more drastically and thus presents more challenges to the HVAC controller. Different building types are used in our experiments, including one-zone building 1 (simplified as \textit{1-zone 1}), four-zone building 1 (\textit{4-zone 1}), four-zone building 2 (\textit{4-zone 2}), four-zone building 3 (\textit{4-zone 3}), five-zone building 1 (\textit{5-zone 1}), seven-zone building 1 (\textit{7-zone 1}). These models are visualized in Figure~\ref{building visual}. In addition, the conditioned air temperature sent from the VAV HVAC system is set to 10 \textcelsius.
	
	The symbols used in the result tables are explained as follows. $\theta_i$ denotes the temperature violation rate in the thermal zone $i$. A$\theta$ and M$\theta$ represent the average temperature violation rate across all zones and the maximum temperature violation rate across all zones, respectively. $\mu_i$ denotes the maximum temperature violation value for zone $i$, measured in \textcelsius. A$\mu$ and M$\mu$ are the average and maximum temperature violation value across all zones, respectively. EP represents the number of training epochs. The symbol ~\CheckedBox ~denotes whether all the temperature violation rates across all zones are less than $5\%$. If it is true, it is marked as $\checkmark$; otherwise, it is $\times$ (which is typically not acceptable for HVAC control).

	\begin{table*}  
		\small
		\vspace{-0.0cm}
		\setlength{\abovecaptionskip}{-0.2cm}
		\setlength{\belowcaptionskip}{-0.0cm}
		\begin{tabular*}{13.3cm}{lllllllllllll}  
			\hline  
			Source building & Target building & $\theta_1$ & $\theta_2$ & $\theta_3$ & $\theta_4$ & $\mu_1$ & $\mu_2$ & $\mu_3$ & $\mu_4$  & \CheckedBox & Cost\\  
			\hline
			\textit{1-zone 1} & \textit{1-zone 1} & 1.62\% & - & - & - & 1.11 & - & - & - & \checkmark & 248.43\\
			\textit{1-zone 1} & \textit{4-zone 2} & 1.88\% & 9.43\% &10.19\% & 14.07\%	& 0.44 &0.97 &1.04 &1.17 & $\times$ & 308.13\\
			\hline  
			
			\hline  
		\end{tabular*}
		\caption{This table shows the experiment that transfers a single-zone DQN model (trained on one-zone building 1) to every zone of four-zone building 2. The high violation rate shows that such a straightforward scheme may not yield good results and more sophisticated methods such as ours are needed.}
		\label{result1-2}
	\end{table*}  
	
	Before reporting the main part of our results, we want to show that simply transferring a well-trained DQN model for a single-zone source building to every zone of a target multi-zone building may not yield good results, as discussed in Section~\ref{sec3-2}. Here as shown in Table~\ref{result1-2}, a DQN model trained for one-zone building 1 works well for itself, but when being transferred directly to every zone of four-zone building 2, there are significant temperature violations. This shows that a more sophisticated approach such as ours is needed. The following sections will show the results of our approach and its comparison with other methods. 
	
	\subsection{Transfer from n-zone to n-zone with different materials and layouts}
	
	In this section, we conduct experiments on building HVAC controller transfer with four-zone buildings that have different materials and layouts. As shown in Figure~\ref{building visual}, four-zone building 1 and four-zone building 2 have different structures, and also different wall materials in each zone with different heat capacities. Table~\ref{result1-3-1} first shows the direct training results on four-zone building 1, and the main transferring results are presented in Table~\ref{result1-3}.
	
	The direct training outcome by baselines and our approach are shown in Table~\ref{result1-3-1}.
	The results include ON-OFF control, Deep Q-network (DQN) control as described in~\cite{wei2017deep} (which assigns an individual DQN model for each zone in the building and trains them for 100 epochs, with one-month data for each epoch), $DQN^*$ (standard deep Q learning method with $m^n$ selections in the last layer~\cite{hester2018deep}), and the direct training result of our method without transferring. Moreover, the DQN method is trained with 50, 100, and 150 training epochs (months), respectively, to show the impact of training time. 
	As shown in the table, all learning-based methods demonstrate significant energy cost reduction over ON-OFF control. 
	$DQN^*$ shows slightly higher cost and violation rate, when compared to DQN after 150 epochs. Our approach with Algorithm \ref{algo1} (i.e., not transferred) achieves the lowest violation rate among all learning-based methods, while providing a low cost.
	
	Table~\ref{result1-3} shows the \textbf{main comparison results} of our transfer learning approach and other baselines on four-zone building 2 and four-zone building 3. ON-OFF, $DQN$ and $DQN^*$ are directly trained on those two buildings. $DQN^*_T$ is a transfer learning approach that transfers a well-trained $DQN^*$ model on four-zone building 1 to the target building (four-zone building 2 or 3). Our approach transfers our trained four-zone building 1 model (last line in Table~\ref{result1-3-1}) to the target building.
	From Table~\ref{result1-3}, we can see that for both four-zone building 2 and 3, with 150 training epochs, $DQN$ and $DQN^*$ provide lower violation rate and cost than ON-OFF control, although $DQN^*$ cannot meet the temperature violation requirement. And the other transfer learning approach $DQN^*_T$ shows very high violation rate. In comparison, our approach achieves extremely low temperature violation rate and a relatively low energy cost without any fine-tuning after transferring (i.e., EP is 0). We may fine tune the controller for 1 epoch (month) after transferring to further reduce the energy cost (i.e., EP is 1), at the expense of slightly higher violation rate (but still meeting the requirement).
	More studies on fine-tuning can be found in Section~\ref{finetune}. Figure~\ref{figure4} (left) also shows the temperature over time for the target four-zone building 2, and we can see that it is kept well within the bounds.

	\begin{table*}  
		\small
		\vspace{-0.0cm}
		\setlength{\abovecaptionskip}{-0.1cm}
		\setlength{\belowcaptionskip}{-0.0cm}
		\begin{tabular*}{12.5cm}{llllllllllllll}  
			\hline  
			Method & Building & EP & $\theta_1$ & $\theta_2$ & $\theta_3$ & $\theta_4$ & $\mu_1$ & $\mu_2$ & $\mu_3$ & $\mu_4$ & \CheckedBox & Cost\\
			\hline
			ON-OFF &  \textit{\textit{4-zone 1}} & \textbf{0} & 0.08\% & 0.08\% & 0.23\% & 0.19\% & 0.01 & 0.03 & 0.08 & 0.08 & \checkmark & 329.56\\
			DQN\cite{wei2017deep} & \textit{4-zone 1} & 50 & 1.21\% & 22.72\% & 9.47\% & 20.66\% & 0.68 & 2.46 & 1.61 & 2.07 & $\times$ &  245.08\\
			DQN\cite{wei2017deep} & \textit{4-zone 1} & 100 & 0.0\% & 0.53\% & 0.05\% & 0.93\% & 0.0 &  0.46 & 0.40 & 1.09 & \checkmark & 292.91\\
			DQN\cite{wei2017deep} & \textit{4-zone 1} & 150 & 0.0\% & 0.95\% & 0.03\% & 1.59\% & 0.0 & 0.52 & 0.17 & 1.17 & \checkmark &  \textbf{278.32}\\
			$DQN^*$ & \textit{4-zone 1} & 150 & 1.74\% & 2.81\% & 1.80\% & 2.76\% & 0.45 & 0.79 & 1.08 & 1.22 & \checkmark & 289.09\\
			Ours & \textit{4-zone 1} & 150 &  \textbf{0.0\%} & \textbf{0.04\%} & \textbf{0.0\%} & \textbf{0.03}\% & 0.0 & 0.33 & 0.0 & 0.11 & \checkmark & 297.42\\
			\hline
			
			\hline  
		\end{tabular*}
		\caption{
			Results of different methods on four-zone building 1. Apart from the ON-OFF control, all others are the training results without transferring. The training model in the last row is used as the transfer model to other buildings in our method.
		}
		\label{result1-3-1}
	\end{table*} 
	
	\begin{table*}  
		\small
		\vspace{-0.0cm}
		\setlength{\abovecaptionskip}{-0.3cm}
		\setlength{\belowcaptionskip}{-0.0cm}
		\begin{tabular*}{12.8cm}{llllllllllllll}  
			\hline  
			Method & Building & EP & $\theta_{1}$  & $\theta_2$  & $\theta_3$  & $\theta_4$  & $\mu_1$  & $\mu_2$  & $\mu_3$  & $\mu_4$  & \CheckedBox & Cost\\
			\hline
			ON-OFF & \textit{4-zone 2} & \textbf{0} & \textbf{0.0\%} & \textbf{0.0\%} & \textbf{0.0\%} & \textbf{0.02\%} & 0.0 & 0.0 & 0.0 & 0.46 & \checkmark & 373.78\\
			DQN\cite{wei2017deep} & \textit{4-zone 2} & 50 & 0.83\% & 49.22\% & 46.75\% & 60.48\% & 0.74 &  2.93 & 3.18 & 3.39 & $\times$ & 258.85\\
			DQN\cite{wei2017deep} & \textit{4-zone 2} & 100 & 0.0\% & 1.67\% & 1.23\% & 3.58\% & 0.0 & 0.92 & 0.77 & 1.62 & \checkmark & 352.13\\
			DQN\cite{wei2017deep} & \textit{4-zone 2} & 150 & 0.0\% & 2.52\% & 1.67\% & 4.84\% & 0.0 & 1.64 & 1.56 & 1.61 & \checkmark & 337.33\\
			$DQN^*$  & \textit{4-zone 2} & 150 & 1.16\% & 2.71\% & 2.17\% & 6.44\%	& 0.61 & 1.11 & 0.77 & 1.11 & $\times$ & 323.72\\
			${DQN^*}_T$ & \textit{4-zone 2} & 0 & 12.35\% & 19.10\% & 10.39\% & 23.59\% & 2.47 & 4.67 & 2.27 & 5.22 & $\times$ & 288.73\\
			Ours & \textit{4-zone 2} & \textbf{0} & \textbf{0.0\%} & \textbf{0.0\%} & \textbf{0.0\%} & \textbf{0.07\%} & 0.0 & 0.0 & 0.0 & 0.88 & \checkmark & 338.45\\
			Ours & \textit{4-zone 2} & \textbf{1} & 0.09\% & 3.44\% & 1.91\% & 4.06\% & 0.33 & 1.04 & 0.96 &  1.35 & \checkmark & \textbf{297.03}\\
			\hline

			ON-OFF & \textit{4-zone 3} & \textbf{0} & \textbf{0.0\%} & \textbf{0.19\%} & \textbf{0.0\%} & \textbf{0.0\%} & 0.0 & 0.02 & 0.0 & 0.0 & \checkmark & 360.74\\
			DQN\cite{wei2017deep} & \textit{4-zone 3} & 50 & 0.68\% & 47.21\% & 44.61\% & 56.19\% & 0.74 & 3.15 & 2.92 & 3.60 & $\times$ & 267.29\\
			DQN\cite{wei2017deep} & \textit{4-zone 3} & 100 & 0.34\% & 2.53\% & 2.21\% & 5.59\% & 0.01 & 1.18 & 0.85 & 1.18 & $\times$ & 342.08\\
			DQN\cite{wei2017deep} & \textit{4-zone 3} &150 & 0.0\% & 1.55\% & 1.68\% & 3.79\% & 0.0 &  1.09 & 1.18 & 1.51 & \checkmark & 334.89\\
			$DQN^*$ & \textit{4-zone 3} & 150 & 7.09\% & 13.85\% & 2.87\% & 2.16\% & 1.26 & 1.48 &  1.42 & 1.01 & $\times$ & 316.93\\
			${DQN^*}_T$ & \textit{4-zone 3} & 0 & 13.31\% & 8.11\% & 3.18\% & 0.66\% & 1.25 & 3.48 & 2.27 & 0.69 & $\times$ & 294.23\\
			Ours & \textit{4-zone 3} & \textbf{0} & \textbf{0.0\%} & \textbf{0.28\%} & \textbf{0.0\%} & \textbf{0.0\%} & 0.0 & 0.37 & 0.0 & 0.0 & \checkmark & 340.40\\
			Ours & \textit{4-zone 3} & \textbf{1} & 0.23\% & 2.74\% & 0.04\% & 0.13\% & 0.34 & 1.73 & 0.12 & 0.31 & \checkmark & \textbf{331.47}\\
			\hline
			
			\hline  
		\end{tabular*}
		\caption{
			Comparison between our approach and other baselines. 
			The top half shows the performance of different controllers on four-zone building 2, including ON-OFF controller, DQN from~\cite{wei2017deep} trained with different number of epochs, the standard Deep Q-learning method ($DQN^*$) and its transferred version from four-zone building 1 ($DQN^*_T$), and our approach transferred from four-zone building 1 (without fine-tuning and with 1 epoch tuning, respectively). We can see that our method achieves the lowest violation rate and very low energy cost after transferring without any further tuning/training. We may fine tune our controller with 1 epoch (month) of training and achieve the lowest cost, at the expense of slightly higher violation rate (but still meeting the requirement). The bottom half shows the similar comparison results for four-zone building 3. 
		}
		\label{result1-3}
	\end{table*}

	\subsection{Transfer from n-zone to m-zone}
	We also study the transfer from an n-zone building to an m-zone building. This is a difficult task because the input and output dimensions are different, presenting significant challenges for DRL network design. Here, we conduct experiments for transferring HVAC controller for four-zone building 1 to five-zone building 1 and seven-zone building 1, and the results are presented in Table~\ref{result5}. 
	For these cases, $DQN^*$ and $DQN^*_T$ cannot provide feasible results as the $m^n$ action space is too large for them, and the violation rate does not go down even after 150 training epochs. 
	$DQN$~\cite{wei2017deep} also leads to high violation rate. In comparison, our approach achieves both low violation rate and low energy cost. Figure~\ref{figure4} (middle and right) shows the temperature over time (kept well within the bounds) for the two target buildings after using our transfer approach.

	\begin{table}  
		\small
		\vspace{-0.0cm}
		\setlength{\abovecaptionskip}{0.1cm}
		\setlength{\belowcaptionskip}{-0.0cm}
		\begin{tabular*}{8.7cm}{lllllllll}  
			\hline  
			Method & Building & EP & A$\theta$ & M$\theta$ & A$\mu$ & M$\mu$ & \CheckedBox & Cost\\  
			\hline
			ON-OFF & \textit{5-zone 1} & \textbf{0} & \textbf{0.45\%} & \textbf{2.2\%} & 0.24 & 1.00 & \checkmark & 373.90 \\
			DQN\cite{wei2017deep} & \textit{5-zone 1} & 50 & 38.65\% & 65.00\% & 2.60 & 3.81 & $\times$ & 263.79 \\
			DQN\cite{wei2017deep} & \textit{5-zone 1} & 100 & 4.13\% & 11.59\% & 4.66 & 1.47 & $\times$ & 326.50 \\
			DQN\cite{wei2017deep} & \textit{5-zone 1} & 150 & 2.86\% & 10.94\% & 0.89 & 1.63 & $\times$ & 323.78 \\
			Ours & \textit{5-zone 1} & \textbf{0} & \textbf{0.47\%} & \textbf{2.34\%} & 0.33 & 1.42 & \checkmark & 339.73\\
			Ours & \textit{5-zone 1} & \textbf{1} & 2.41\% & 4.48\% & 1.02 & 1.64 & \checkmark & \textbf{323.26}\\
			\hline
			
			ON-OFF & \textit{7-zone 1} & \textbf{0} & \textbf{0.37\%} & \textbf{2.61\%} & 0.04 & 0.30 & \checkmark & 392.56 \\
			DQN\cite{wei2017deep} & \textit{7-zone 1} & 50 & 28.14\% & 54.28\% & 2.76 & 3.06 & $\times$ & 248.38 \\
			DQN\cite{wei2017deep} & \textit{7-zone 1} & 100 & 5.19\% & 18.91\% & 1.12 & 1.69 & $\times$ & 277.87 \\
			DQN\cite{wei2017deep} & \textit{7-zone 1} & 150 & 4.48\% & 18.34\% & 1.22 & 1.98 & $\times$ & 284.51 \\
			Ours & \textit{7-zone 1} & \textbf{0} & \textbf{0.42\%} & \textbf{2.79\%} & 0.10 & 0.43 & \checkmark & 332.07\\
			Ours & \textit{7-zone 1} & \textbf{1} & \textbf{0.77\%} & \textbf{1.16\%} & 0.77 & 1.21 & \checkmark & \textbf{329.81}\\
			\hline
			
			\hline  
		\end{tabular*}
		\caption{Comparison of our approach and baselines on five-zone building 1 and seven-zone building 1.}
		\label{result5}
	\end{table}

	\begin{figure*}[htbp]
		\centering
		\vspace{-0.0cm}
		\setlength{\abovecaptionskip}{-0.2cm}
		\setlength{\belowcaptionskip}{-0.2cm}
		\includegraphics[width=18cm]{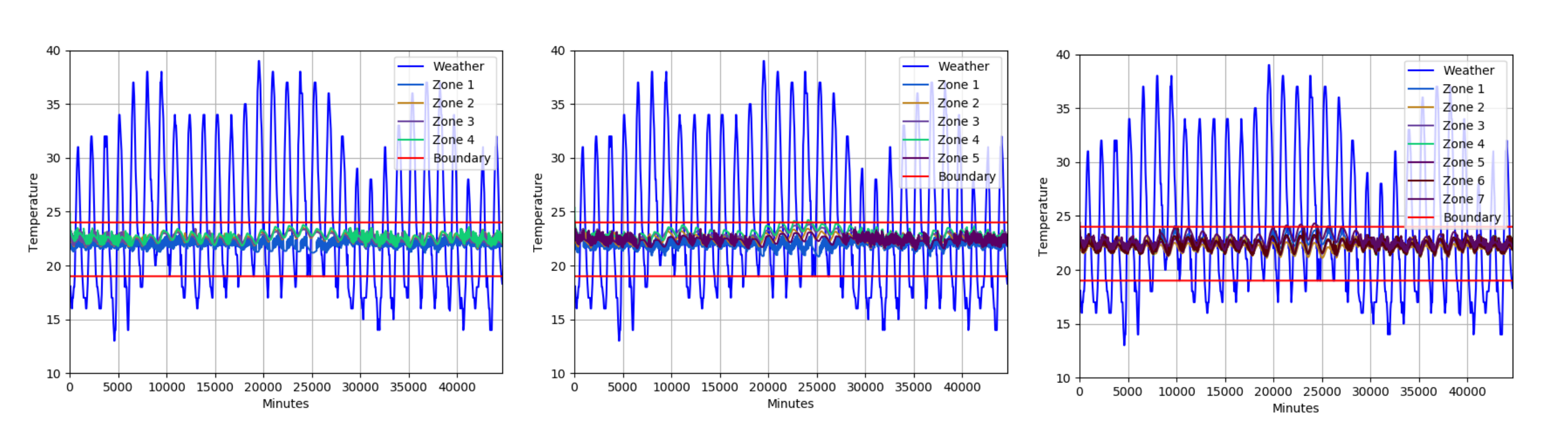}
		\caption{Temperature of four-zone building 2 (left), 5-zone building 1 (middle), and 7-zone building 1 (right) after transfer.}
		\label{figure4}
	\end{figure*}
	
	\subsection{Transfer from n-zone to n-zone with different HVAC equipment}
	In some cases, the target building may have different HVAC equipment (or a building may have its equipment upgraded). 
	The new HVAC equipment may be more powerful or have a different number of control levels, making the original controller not as effective. 
	In such cases, our transfer learning approach provides an effective solution. 
	Here we conduct experiments on transferring our controller for the original HVAC equipment (denote as AC 1, which has two control levels and used in all other experiments) to the same building with new HVAC equipment (denoted as AC2, which has five control levels; and AC3, which has double max airflow rate and double air conditioner power compared to AC1). The experimental results are shown in Table~\ref{result3}. We can see that our approach provides zero violation rate after transferring, and the energy cost can be further reduced with the fine tuning process.

	\begin{table}  
		\small
		\vspace{-0.3cm}
		\setlength{\abovecaptionskip}{-0.3cm}
		\setlength{\belowcaptionskip}{-0.0cm}
		\begin{tabular*}{8.7cm}{llllllllll}  
			\hline  
			Method & AC & EP & $A\theta$ & M$\theta$ & A$\mu$ & M$\mu$ & \CheckedBox & Cost\\  
			\hline
			ON-OFF & AC 2 & \textbf{0} & \textbf{0.15\%} & \textbf{0.23\%} & 0.05 & 0.08 & \checkmark & 329.56\\
			DQN\cite{wei2017deep} & AC 2 & 50 & 20.28\% & 35.56\% & 1.73 & 2.66 & $\times$ &  229.41\\
			DQN\cite{wei2017deep} & AC 2 & 100 & 1.25\% & 2.69\% & 0.61 & 1.20 & \checkmark &  270.93\\
			DQN\cite{wei2017deep} & AC 2 & 150 & 1.49\% & 2.87\% & 0.60 & 1.02 & \checkmark &  263.92\\
			Ours & AC 2 & \textbf{0} & \textbf{0.0\%} & \textbf{0.0\%} & 0.0 & 0.0 & \checkmark & 303.37\\
			Ours & AC 2 & \textbf{1} & 2.06\% & 4.20\% & 0.97 & 1.30 & \checkmark & \textbf{262.23}\\
			\hline
			ON-OFF & AC 3 & \textbf{0} & \textbf{0.01\%} & \textbf{0.05\%} & 0.22 & 0.88 & \checkmark & 317.53\\
			DQN\cite{wei2017deep} & AC 3 & 50 & 2.85\% & 3.76\% & 1.37 & 1.90 & \checkmark & 321.03\\
			DQN\cite{wei2017deep} & AC 3 & 100 & 0.69\% & 1.20\% & 0.53 & 0.99 & \checkmark & \textbf{265.46}\\
			DQN\cite{wei2017deep} & AC 3 & 150 & 0.62\% & 1.07\% & 0.47 & 0.65 & \checkmark & 266.86\\
			Ours & AC 3 & \textbf{0} & \textbf{0.0\%} & \textbf{0.0\%} & 0.0 & 0.0 & \checkmark & 316.16\\
			Ours & AC 3 & \textbf{1} & 0.84\% & 1.42\% & 0.54 & 0.78 & \checkmark & 269.24\\
			\hline
			
			\hline  
		\end{tabular*}
		\caption{Comparison under different HVAC equipment.}
		\label{result3}
	\end{table}

	\subsection{Fine-tuning study}
	\label{finetune}
\begin{figure}[htbp]
		\vspace{-0.0cm}
		\setlength{\abovecaptionskip}{-0.0cm}
		\setlength{\belowcaptionskip}{-0.1cm}
		\centering
		
		\includegraphics[width=8.5cm]{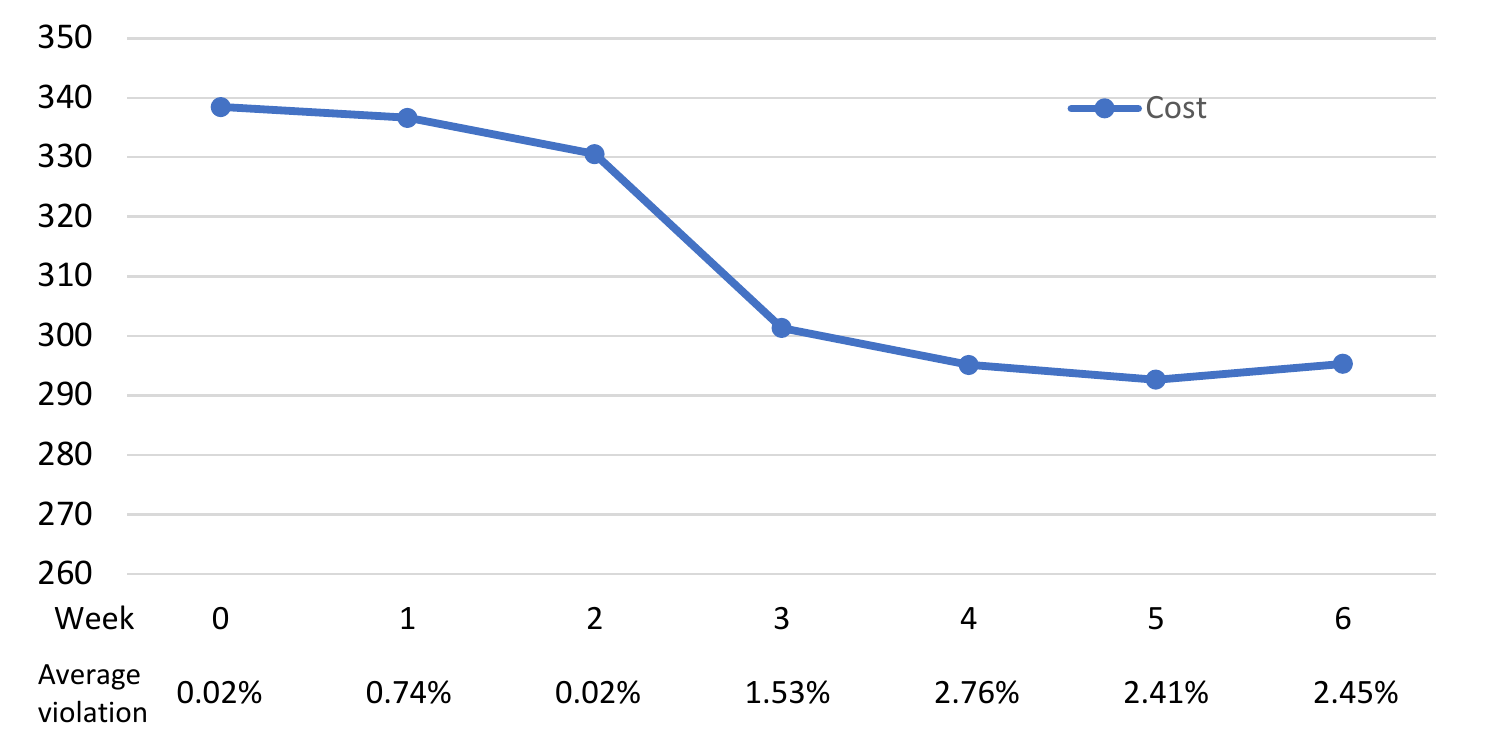}
		\caption{Fine-tuning results of our approach for four-zone building 2. Our approach can significantly reduce energy cost after fine-tuning for 3 weeks, while keeping the temperature violation rate at a low level.}
		\label{figure6}
	\end{figure}

	After transferring, although our method has already gained a great performance without fine-tuning, further training is still worth considering because it may provide even lower energy cost. We record the change of cost and violation rate when fine-tuning our method transferred from four-zone building 1 to four-zone building 2. The results are shown in Figure~\ref{figure6}.

	\subsection{Discussion}
	
	\subsubsection{Transfer from n-zone to n-zone with different weather}
\begin{figure}[htbp]
		\centering
		\vspace{-0.3cm}
		\setlength{\abovecaptionskip}{-0.0cm}
		\setlength{\belowcaptionskip}{-0.6cm}
		\includegraphics[width=7cm]{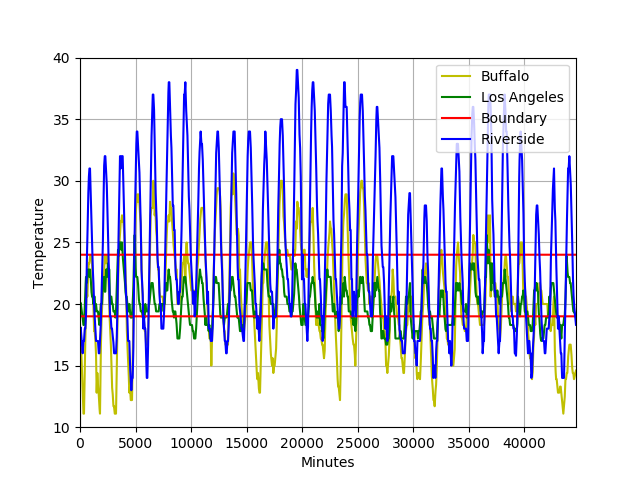}
		\caption{The visualization of different weathers. The yellow line is the Buffalo weather, the green line is the LA weather, the blue line is the Riverside weather, and the red lines are the comfortable temperature boundary.}
		\label{weather}
	\end{figure}

	As presented in~\cite{lissa2020transfer}, the Q-learning controller with weather that has a larger temperature range and variance is easy to be transferred into the environment with the weather that has a smaller temperature range and variance, but it is much harder in the opposite direction. 
	This conclusion is similar to what we observed for our approach.  
	We tested the weather from Riverside, Buffalo, and Los Angeles, which is shown in Figure~\ref{weather}. The results show that our approach can easily be transferred from large range and high variance weather (Riverside) to small range and low variance weather (Buffalo and Los Angeles(LA)), but not vice versa. 
	Fortunately, the transferring for a new building is still not affected, because our approach can use the building models in the same region or obtain the weather data in that region and create a simulated model for transferring.

	\begin{table}  
		\small
		\vspace{-0.0cm}
		\setlength{\abovecaptionskip}{-0.5cm}
		\setlength{\belowcaptionskip}{-0.0cm}
		\begin{tabular*}{8.6cm}{llllllllll}  
			\hline  
			Building & Source & Target & EP & A$\theta$ & M$\theta$ & \CheckedBox & Cost\\  
			\hline  
			\textit{4-zone 1} & LA & LA & 150 & 0.68\% & 1.71\% & \checkmark & 82.01\\
			\textit{4-zone 1} & Buffalo & Buffalo & 150 & 0.64\% & 1.14\% & \checkmark & 101.79\\
			\textit{4-zone 1} & Riverside & Riverside & 150 &  0.02\% & 0.04\% & \checkmark & 297.42\\
			
			\textit{4-zone 1} & Riverside & LA & 0 & 0.0\% & 0.0\% &  \checkmark &  105.17\\
			\textit{4-zone 1} & Riverside & Buffalo & 0 & 0.0\% & 0.0\% &  \checkmark & 134.28\\
			\textit{4-zone 1} & LA & Riverside & 0 & 71.77\% & 89.34\% &  $\times$ & 158.06\\
			\textit{4-zone 1} & Buffalo & Riverside & 0 & 54.92\% & 81.89\%  & $\times$ & 180.20\\
			\hline
			
			\hline  
		\end{tabular*}
		\caption{Transferring between different weathers.}
		\label{result2}
	\end{table} 
	
	\subsubsection{Different settings for ON-OFF control}
	\label{ON-OFF}
	Our back-end network (inverse building network) is learned from the dataset collected by an ON-OFF control with low temperature violation rate. In practice, it is flexible to determine the actual temperature boundaries for ON-OFF control. For instance, the operator may set the temperature bound of ON-OFF control to be within the human comfortable temperature boundary (what we use for our method) or just the same as the human comfortable temperature boundary, or even a little out of boundary to save energy cost. Thus, we tested the performance of our method by collecting data under different ON-OFF boundary settings.  Results in Table~\ref{boundary_test} shows that with different boundary settings, supervised learning can stably learn from building-specific behaviors.
	
	\begin{table}  
		\vspace{-0.0cm}
		\small
		\setlength{\abovecaptionskip}{-0.0cm}
		\setlength{\belowcaptionskip}{-0.0cm}
		\begin{tabular*}{7cm}{llllllll}  
			\hline  
			Method & Upper-Bound & EP & A$\theta$ & M$\theta$ & Cost\\
			\hline
			ON-OFF & 23 & 0 & 0.01\% & 0.02\% & 373.78\\
			ON-OFF & 24 & 0 & 61.45\% & 73.69\% & 256.46\\
			ON-OFF & 25 & 0 & 98.56\% & 99.99\% & 208.79\\
			\hline
			Ours & 23 & 0 & 0.02\% & 0.07\% & 338.45\\
			Ours & 24 & 0 & 0.02\% & 0.07\% & 338.08\\
			Ours & 25 & 0 & 0.02\% & 0.07\% & 338.08\\
			\hline
			
			\hline  
		\end{tabular*}
		\caption{Results of testing using different boundary.}
		\label{boundary_test}
	\end{table}

	\section{Conclusion}
	\label{conclusion}
	In this paper, we present a novel transfer learning approach that decomposes the design of the neural network based HVAC controller into two sub-networks: a building-agnostic front-end network that can be directly transferred, and a building-specific back-end network that can be efficiently trained with offline supervise learning. Our approach successfully transfers the DRL-based building HVAC controller from source buildings to target buildings that can have a different number of thermal zones, different materials and layouts, different HVAC equipment, and even under different weather conditions in certain cases.

	\begin{acks}
		
		We gratefully acknowledge the support from Department of Energy (DOE) award DE-EE0009150 and National Science Foundation (NSF) award 1834701.
	\end{acks}

	\bibliographystyle{ACM-Reference-Format}
	\bibliography{sample-base, yixuan}

\end{document}